\documentclass[12pt,epsf]{article}
\usepackage{amssymb,amsmath}
\usepackage{graphicx}
\usepackage{setspace}
\usepackage{indentfirst} 

\newcommand{\be}{\begin{equation}}
\newcommand{\ee}{\end{equation}}
\newcommand{\bea}{\begin{eqnarray}}
\newcommand{\eea}{\end{eqnarray}}
\newcommand{\bear}{\begin{eqnarray}}
\newcommand{\eear}{\end{eqnarray}}
\newcommand{\beas}{\begin{eqnarray*}}
\newcommand{\eeas}{\end{eqnarray*}}
\newcommand{\ba}{\begin{array}}
\newcommand{\ea}{\end{array}}

\newcommand{\bali}{\begin{aligned}}
\newcommand{\eali}{\end{aligned}}

\makeatletter

\newcommand{\Rmnum}[1]{\expandafter\@slowromancap\romannumeral #1@}
\makeatother

\renewcommand{\a}{\alpha}
\renewcommand{\b}{\beta}
\newcommand{\De}{\Delta}
\newcommand{\m}{\mu}

\newcommand{\f}{\phi} \newcommand{\F}{\Phi}

\newcommand{\tr}{\mathrm{Tr}}

\newcommand{\ov}{\overline}
\newcommand{\Hcal}{\mathcal{H}}
\newcommand{\Hcalb}{\ov{\mathcal{H}}}
\newcommand{\da}{\dagger}

\newcommand{\nbox}{{\,\lower0.9pt\vbox{\hrule \hbox{\vrule height 0.2 cm \hskip 0.19 cm \vrule height 0.2 cm}\hrule}\,}}

\def\href#1#2{#2}

\begin{document}
\begin{titlepage}
\hfill
\vbox{
    \halign{#\hfil         \cr
           } 
      }  
\vspace*{20mm}
\begin{center}
{\Large \bf Bulk Cluster Decomposition in AdS/CFT and A No-Go Theorem for Correlators in Microstates of Extremal Black Holes}

\vspace*{16mm}
Byungwoo Kang
\vspace*{1cm}

{
Department of Physics,
Stanford University\\
Stanford, CA 94305\\
\emph{redcrux8@stanford.edu}}

\vspace*{1cm}
\end{center}
\begin{abstract}
Applying the thermo-field double formalism to extremal black holes in AdS with a macroscopic horizon, we show that (1) there exists a natural basis for the degenerate microstates of an extremal black hole, and (2) cluster decomposition in the bulk implies that all correlators are exactly the same for every microstate of the extremal black hole. The latter statement can be interpreted in two ways. First, at the fully non-perturbative level of AdS/CFT at finite $N$, it means that cluster decomposition does not hold in the bulk. This may be viewed as a sharp manifestation of the bulk non-locality at finite $N$. Second, at the level of the perturbation theory in $1/N$, in which case we expect the bulk cluster decomposition, no measurement of either boundary operators or bulk field operators can distinguish the different microstates. The latter interpretation may exclude some versions of the fuzzball conjecture that assert that different microstates of a black hole are realized in the bulk as different metric and field configurations.
\end{abstract}

\end{titlepage}

\vskip 1cm

\section{Introduction and summary}
Although we have certain microscopic understanding of the black hole entropy since the pioneering work of Strominger and Vafa \cite{strovafa}, we still do not have good understanding of the detailed properties of the black hole microstates. In this paper, we take a step toward better understanding of the microstates of extremal black holes in AdS. Specifically, we show that, if cluster decomposition holds in the bulk spactime, the values of all correlators are identical for every microstate of an extremal black hole in AdS. The observables in consideration could be in general a product of an arbitrary number of operators, not necessarily of the same kind and with no restriction on insertion points (e.g. time-like seperation between probe operators is possible). In particular, the operators could be either boundary field theory operators or ``bulk field'' operators constructed from dual boundary operators, which we will explain in detail later.

The main strategy of the argument is as follows. We consider a family of black holes in AdS either at fixed charges and/or angular momenta (canonical ensemble) or fixed chemical potentials (grandcanonical ensemble) parametrized by temperature. We demand that they have a macroscopic horizon at any temperature and in the zero temperature limit and that the black hole solution is globally stable (i.e. the dominant saddle) in the zero temperature limit. In fact, it has been shown that, for cetain ranges of charges or chemical potentials, (near-)extremal charged and/or rotating black holes with a macroscopic horizon are the dominant saddles in the zero temperature limit \cite{emp1,emp2,calda, louko, peca, rota1, rota2}.\footnote{In certain cases, it has been only shown that they are locally stable due to difficulties in finding competing saddles in a given ensemble. See for instance \cite{calda} for the discussion of this point. In case of rotating black holes, we also demand that they be stable against superradiance. Nice discussions of superradiance and its relation to the instability of rotating black holes can be found in \cite{kodama, rota1, murata}.} The maximally extended solutions of these black holes correspond to thermo-field double states in the boundary theory at fixed charges or chemical potentials \cite{malda01}. 

As the temperature is decreased to zero, the exterior of the black hole develops an infinite throat in the near-horizon region. This implies that any two points in the opposite exterior regions, at fixed radial coordinates of a standard metric for the family of the black holes, become infinite far away from each other in the zero temperature limit. If cluster decomposition holds in the bulk, then, in the zero temperature limit, the correlator of a product of operators on one exterior region and those on the other exterior region should factorize into a product of the correlator of the operators in one exterior region and its counterpart in the other region. Then, using the formalism of thermo-field double, one can show that any one-sided correlators evaluated in any microstate of the extremal black hole is equal to that evaluated in the zero-tempreature ensemble of the microstates.

The paper has the following structure. In Section 2, we define thermo-field double states for a family of charged and/or rotating black holes in AdS in canonical or grandcanonical ensembles and explain how to deal with subtleties in doing so when there are degenerate vacua. In Section 3, we review the thermo-field double formalism with an emphasis on the properties crucial in establishing the argument. In Section 4, we present the argument, discuss its implications and subtleties, and future research directions. In particular, we discuss how this result may exclude certain versions of the fuzzball conjecture.\footnote{For reviews of the fuzzball proposal, see \cite{fuzz1,fuzz2,fuzz3} and references therein.}   

\section{Thermo-field double for charged and/or rotating black holes}   

The thermo-field double state of two copies of a (strongly coupled) CFT without any charge or chemical potential,
\be
\label{neutral}
|\Psi_0 \rangle = {1 \over \sqrt{Z(\b)}} \sum_{E_i} e^{- \b E_i} |E_i \rangle \otimes |\ov{E_i} \rangle , \text{ where $|E_i \rangle \in \Hcal$ and $|\ov{E_i} \rangle \in \Hcalb$},
\ee
desribes a large AdS Schwarzschild black hole for temperatures higher than the Hawking-Page temperature, or two seperate thermal AdS spaces for lower temperatures. \cite{malda01, hp}. To describe an AdS black hole with charges (angular momenta will also be called charges from here on, unless otherwise mentioned), we need to consider states that have fixed charges or introduce chemical potentials. The corresponding thermo-field double state is
\be
\label{canonical}
|\Psi \rangle = {1 \over \sqrt{Z(\b)}} \sum_{E_i} e^{- {\b \over 2} E_i} |E_i, q_1, q_2, ... q_n \rangle \otimes |\ov{E_i, q_1, q_2, ... q_n} \rangle,
\ee
in case of canonical ensemble, or
\be
\label{canonical}
|\Psi \rangle = {1 \over \sqrt{Z(\b,\m)}} \sum_{E_i, q_1, q_2, ... q_n} e^{- {\b \over 2}(E_i - \sum_{j}\m_j q_j)} |E_i, q_1, q_2, ... q_n \rangle \otimes |\ov{E_i, q_1, q_2, ... q_n} \rangle,
\ee
in case of grandcanonical ensemble, where $q_i$'s are the charges and $\m_i$'s are the corresponding chemical potentials. The argument presented below applies for both canonical and grandcanonical ensemble; in case of grandcanonical ensemble, we may simply regard $H - \sum_{j}\m_j Q_j$ as the new Hamiltonian, where $H$ is the original Hamiltonian and $Q_j$'s are the charge operators. In fact, in case of the AdS Kerr black hole, one can easily see that using this new Hamiltonian corresponds to ``time''-evolution with respect to the Killing vector whose Killing horizon defines the event horizon. Therefore, without loss of generality, and for notational convenience, we simplify the notation of $|\Psi \rangle$ as
\be
\label{psi}
|\Psi \rangle = {1 \over \sqrt{Z}} \sum_{i} e^{ -{\b \over 2} E_i} |i \rangle \otimes |\ov{i} \rangle,
\ee
where the sum is over all states of varying energy and fixed charges, or over all states of varying energy and charges in case of canonical ensemble.

Now, there is an ambiguity in defining $|\Psi \rangle$ when there are degenerate vacua: which basis for the degenerate vacua should appear in $|\Psi \rangle$? In fact, since the zero-temperature limit of the family of the black holes described by $|\Psi \rangle$ is assumed to have a macroscopic horizon, there should be a large ground state degeneracy in the system.\footnote{In case of non-supersymmetric AdS extremal black holes, the ground state degeneracy might be lifted by energy spacings of order $E_{gap}e^{-S_0}$, where $E_{gap}$ is the mass gap of the black hole \cite{presk} and $S_0$ is the Bekenstein-Hawking entropy of the extremal black hole (see, for example, \cite{page} for discussion of this point). If that is true, we cannot lower the temperature strictly to zero in (\ref{bulkmain}) and (\ref{main}) since otherwise only the unique ground state would survive in the zero temperature limit; we should lower the temperature only up to some point so that we retain most of the states that contribute to $S_0$ in the ensemble in (\ref{n}). This would cause a certain amount of error to (\ref{result}). However, even when the ground state degeneracy is expected to be exact as in the case of supersymmetric black holes, the bulk cluster decomposition should break down non-perturbatively in $N$ as we discuss below. Therefore, as long as the error due to the non-exact ground state degeneracy is parametrically smaller than powers of $1/N$, which we believe to be true at least for a large class of operators (including those dual to massive bulk fields), the conclusions in the abstract and below would still be valid. Similarly, it may turn out that semiclassical approximation is not reliable when the temperature is lower than the mass gap, and thus, if we are to use semiclassical concepts like cluster decomposition, we should not lower the temperature below the mass gap in (\ref{bulkmain}) and (\ref{main}). Even if that is the case, the same remarks as above will hold for the error due to the mass gap.} It turns out that, as we will see below in (\ref{result}), the right basis to choose is the basis in which the off-diagonal matrix elements of all Hermitian local operators in the boundary field theory vanish in the large $N$ limit. This basis is unique in general, and its basis elements satisfy cluster decomposition in the boundary field theory if the spatial conformal boundary is not compact (for example, if the spatial conformal boundary is $S^3$, cluster decomposition in the boundary field theory cannot be sharply formulated since the distance between any two points on the sphere is finite).\footnote{We emphasize that cluster decomposition in the boundary field theory is preserved and should not be confused with cluster decomposition in the bulk which must be violated at finite $N$ as we show below.} A thorough discussion of the existence and properties of this basis when the conformal boundary is flat is presented in Section 19.1 of \cite{wein2}. The existence of this basis when the conformal boundary is compact is non-trivial at least from the point of view of \cite{wein2} and seems closely related to the large $N$ limit.            

\section{Review of the thermo-field double formalism}
In this section, we review the thermo-field double formalism with an emphasis on the aspects relevant for the main argument.\footnote{Our presentation closely follows the original paper on the thermo-field double formalism by Takahashi and Umezawa \cite{thermo}.} First, given a Hamiltonian $H$ with the properties
\bea
\bali
H |n\rangle = E_n |n\rangle \\
\langle n | m \rangle = \delta_{nm},
\eali
\eea
we define $\ov{H}$ such that 
\bea
\bali
\ov{H} |\ov{n}\rangle = E_n |\ov{n}\rangle \\
\langle \ov {n} | \ov{m} \rangle = \delta_{nm}.
\eali
\eea  
If there are more quantum numbers like conserved charges, we may similarly define their counterparts in $\mathcal{H}$, but we will suppress other quantum numbers. Now, for any operator $A$ in $\Hcal$, define $\ov{A}$ in $\Hcalb$ such that
\be
\langle i | A | j \rangle = \langle \ov{j} | \ov{A}^{\da} | \ov {i} \rangle, \; \; \forall i, j.
\ee  
Also, define
\bea
A(t) \equiv e^{iHt} A e^{-iHt}, \\
\ov{A}(t) \equiv e^{-i\ov{H}t}\ov{A}e^{i\ov{H}t}.	
\eea
Note that $\ov{A}(t)$ evolves in time with the ``Hamiltonian'' $-\ov{H}$. These definitions ensure that
\be
\langle i | A(t) | j \rangle = \langle \ov{j} | \ov{A}^{\da}(t) | \ov{i} \rangle, \; \; \forall i,j \text{ and } t.
\ee 
Using the above definitions, it is straightforward to show that
\be
\label{lemma1}
\ov{A_1 A_2 \cdots A_n} = \ov{A_1} \; \ov{A_2} \; \cdots \; \ov{A_n},
\ee
and
\be
\ov{\a_1 A_1 + \a_2 A_2} = \a_1^{\ast} \ov{A_1} + \a_2^{\ast} \ov{A_2},
\ee
where $\a_1$ and $\a_2$ are complex c-numbers. A crucial property of the thermo-field double state $|\Psi\rangle$ defined in (\ref{psi}) is that a two-sided correlator in that state can be obtained from analytic continuation of a one-sided correlator. In particular,
\bea 
\langle \Psi | A(t) B(t'+i\b /2) | \Psi \rangle
&=& {1 \over Z}\sum_{i,j} e^{-\b E_i} \langle i | A | j \rangle \langle j | B | i \rangle e^{-i(E_j-E_i)(t-t'-i\b/2)} \nonumber \\
&=& {1 \over Z}\sum_{i,j} e^{-{\b \over 2}(E_i+E_j)}\langle i | A | j \rangle \langle \ov{i} | \ov{B}^{\da} | \ov{j} \rangle e^{-i(E_j-E_i)(t-t')} \nonumber \\
&=& {1 \over Z} \sum_{i,j} e^{-{\b \over 2}(E_i+E_j)}\langle i | A(t) | j \rangle \langle \ov{i} | \ov{B}^{\da} (t') | \ov{j} \rangle \nonumber \\
&=& \langle \Psi | A(t) \ov{B}^{\da} (t') | \Psi \rangle.
\eea
Finally, we present some expressions that will be essential in the below argument:
\bea
\label{A}
\langle \Psi | A | \Psi \rangle &=& {1 \over Z} \sum_i e^{-\b E_i} A_{ii}, \text{ where $A_{ii} \equiv \langle i | A | i \rangle$}, \\
\label{Abar}
\langle \Psi | \ov{A} | \Psi \rangle &=& {1 \over Z} \sum_i e^{-\b E_i} A_{ii}^{\ast}, \\
\label{AAbar}
\langle \Psi | A\, \ov{A} | \Psi \rangle &=& {1 \over Z} \sum_{i,j} e^{-{\b \over 2}(E_i + E_j)} \langle i|A|j \rangle \langle \ov{i}|\ov{A}|\ov{j} \rangle \nonumber\\ 
&=& {1 \over Z} \sum_{i,j} e^{-{\b \over 2}(E_i + E_j)} \langle i|A|j \rangle \langle i|A|j \rangle^{\ast} \nonumber\\
&=& {1 \over Z} \sum_{i,j} e^{-{\b \over 2}(E_i + E_j)} |A_{ij}|^2. 
\eea

\section{The main argument and discussion}
Throughout this section, we consider the case of possibly large but finite $N$, in which case the black hole entropy is finite\footnote{In case of the planar black hole, we may assume that its horizon is toroidally compactified to ensure the finiteness of its entropy.} and therefore there are no potential subtleties with replacing the sum by an integral in the thermo-field double representation of $|\Psi\rangle$. Let 
\begin{align}
\F^{\Rmnum{1}} &\equiv \f_1^{\Rmnum{1}}(r_1,x_1)\f_2^{\Rmnum{1}}(r_2,x_2) \cdots \f_n^{\Rmnum{1}}(r_n,x_n),\\
\F^{\Rmnum{2}} &\equiv \f_1^{\Rmnum{2}}(r_1,x_1)\f_2^{\Rmnum{2}}(r_2,x_2) \cdots \f_n^{\Rmnum{2}}(r_n,x_n),
\end{align}
where $\f_i^{\Rmnum{1} (\Rmnum{2})}$ is a bulk field in the exterior region $\Rmnum{1} (\Rmnum{2})$. The coordinate r is the radial coordinate and x collectively denotes all the other coordinates. We use Schwarzschild-like coordinates here and both exterior regions are covered by them in a standard way.\footnote{For rotating black holes, we may choose to use Boyer-Lindquist-like coordinates or other similar coordinates. See \cite{krish1,krish2,chand} for nice discussions on coordinate systems in rotating black holes.} In particular, the two points that have the same coordinates but in different exterior regions are related to each other on the Penrose diagram by reflection through the intersection of the bifurcating Killing horizons. The boundary field theory operator dual to $\f_i^{\Rmnum{1}}$ is denoted by $O_i$. Then, due to the thermo-field double nature of the two-sided black hole, the operator dual to $\f_i^{\Rmnum{2}}$ is given by $\ov{O_i}$ \cite{israel,malda01,infall}. 

Now, consider the bulk correlator $\langle \F^{\Rmnum{1}}\F^{\Rmnum{2}} \rangle_{\text{bulk}}$. We first fix the coordinates $r_i$ and $x_i$ and take the zero-temperature limit. $r_i$'s are taken sufficiently large so that the bulk field operators are located always outside the horizon as we decrease the temperature. As the temperature goes to zero, the metric will change correspondingly and in particular develop an infinite throat in the near-horizon region \cite{rozali,horo,reall,rangam}. In case of the AdS Reissner-Nordstrom black hole, the fact that its blackening factor has a local quadratic extremum implies that there exist real geodesics connecting any two points on the different exterior regions (in contrast to the AdS Schwarzschild black hole \cite{shenker}) and that their lengths diverge in the zero temperature limit \cite{rozali}. In case of rotating black holes, because of fewer rotational symmetries there are fewer Killing conserved quantities, and therefore it is in general harder to analyze the geodesics.\footnote{In certain cases such as four and five dimensional AdS Kerr black holes, there are additional conserved quantities for such geodesics not explicitly related to the spacetime symmetry, which make it possible to solve the geodesic equations in a manner similar to the case of non-rotating black holes \cite{saka,luci}.} By analogy with the Reissner-Nordstrom black hole, however, similar conclusions are expected to hold for rotating black holes as well. We will assume that the divergence of the length of any real geodesic connecting the two exterior regions in the zero temperature limit implies cluster decomposition between these exterior regions. An important caveat here is that, in curved spacetimes, cluster decomposition may not be governed by real geodesic distances but by complex geodesic distances, as shown explicitly in case of the AdS Schwarzschild black hole in spacetime dimensions higher than three \cite{shenker,liu}. However, a general analysis of complex geodesics and their contributions to bulk correlators in general charged and/or rotating black holes is very complicated, and we will not concern ourselves about this subtlety in this paper. Then, by the (conjectured) cluster decomposition between the two exterior regions, 
\be
\label{bulkmain}
\lim_{\b \to \infty} \langle \F^{\Rmnum{1}}\F^{\Rmnum{2}} \rangle_{\text{bulk}} = \lim_{\b \to \infty} \langle \F^{\Rmnum{1}} \rangle_{\text{bulk}} \lim_{\b \to \infty} \langle \F^{\Rmnum{2}} \rangle_{\text{bulk}}.    
\ee
From this and the AdS/CFT dictionary, we find that\footnote{One might be worried about the order of the limits in the first line of (\ref{main}). Alternatively, one may explicitly impose a UV-cutoff instead of taking the $r_i \to \infty$ limit, and take the zero temperature limit. This is equivalent to the order of the limits used in (\ref{main}). The other order of the limits is also valid, essentially by definition. Therefore, the two limits commute.}
\be
\label{main}
\bali
\langle \Psi | \prod_i O_i(x_i)\prod_i \ov{O_i}(x_i) | \Psi \rangle_{\b = \infty} 
&= \lim_{r_i \to \infty} \prod_i r_i^{2\De_i}  \langle \F^{\Rmnum{1}}\F^{\Rmnum{2}} \rangle_{\b=\infty} \\
&= \lim_{r_i \to \infty} \prod_i r_i^{2\De_i}  \langle \F^{\Rmnum{1}} \rangle_{\b=\infty} \langle \F^{\Rmnum{2}} \rangle_{\b=\infty} \\
&= \langle \Psi | \prod_i O_i(x_i) | \Psi \rangle_{\b = \infty} \langle \Psi | \prod_i \ov{O_i}(x_i) | \Psi \rangle_{\b = \infty} \\
&= \langle \Psi | \prod_i O_i(x_i) | \Psi \rangle_{\b = \infty} \langle \Psi | \ov{\prod_i O_i(x_i)} | \Psi \rangle_{\b = \infty}.
\eali
\ee
In the above equation, $\prod_i O_i(x_i) = O_1(x_1)\cdots O_n(x_n)$ and similarly for $\ov{O_i}$'s,\footnote{Note that the ordering of the operators should be specified since they do not necessarily commute with each other, especially when they are time-like seperated.} and the subscript ``$\b=\infty$'' of $|\Psi\rangle$ means the zero-temperature limit of $|\Psi\rangle$. Also, $\langle \cdots \rangle_{\b=\infty} \equiv \lim_{\b \to \infty} \langle \cdots \rangle_{\text{bulk}}$, and we used (\ref{lemma1}) in going from the third to fourth lines. 

Applying the zero temperature limit of (\ref{A}), (\ref{Abar}), and (\ref{AAbar}) to (\ref{main}), with $A = \prod_i O_i(x_i)$, we obtain that
\be
{1\over n}\sum_{i,j} |A_{ij}|^2 = \left|{1\over n}\sum_{i} A_{ii}\right|^2 \label{n}
\ee
\begin{gather}
\Rightarrow \left<|A|^2\right> + {1\over n}\sum_{i\not= j} |A_{ij}|^2 = |\left<A\right>|^2 \label{average}\\
\Rightarrow \left<|A-\left<A\right>|^2\right> = - {1\over n}\sum_{i\not= j} |A_{ij}|^2 \label{identity}\\
\Rightarrow \left<|A-\left<A\right>|^2\right> = 0, \; \text{   and   } \; |A_{ij}|=0, \; \forall \; i\not= j. \label{result}
\end{gather}
In (\ref{n}), $n$ is the number of the degenerate ground states, and the summation indices label the degenerate ground states. In (\ref{average}), $\left<\cdots\right> \equiv 1/n\sum_{i}\cdots$ is the ensemble average over the degenerate ground states. (\ref{result}) says that any (local) boundary correlator in any microstate of the extremal black hole is equal to that in the thermo-field ensemble of the extremal black hole microstates and that all off-diagonal matrix elements between different microstates vanish, even at finite $N$ and finite volume.\footnote{It is shown in Section 19.1 of \cite{wein2} that off-diagonal matrix elements of local operators between different vacua vanish in the infinite volume limit.} However, of course, this cannot be true at the fully non-perturbative level at finite $N$, since different states cannot have exactly the same correlators for all operators. Therefore, we conclude that, at finite $N$, the bulk clsuter decomposition must be violated. On the other hand, in the bulk perturbation theory in $1/N$, we expect cluster decomposition to hold at the level of supergravity and quite likely also at the level of perturbative string theory. This suggests that the violation of the bulk cluster decomposition is non-perturbative. Still, it is possible that, for some subtle reason, cluster decomposition is violated at a pertubative level. In particular, such effective field theoretical or perturbative string calculations may go wrong, for instance when the number of operators inside the correlator scales with some power of $N$ \cite{infall,arkani}. This isssue deserves a further investigation.

Another approach to deriving (\ref{main}) uses the Ryu-Takayanagi holographic entanglement formula \cite{RT1,RT2,HRT}. Let $I(A:B)$ be the mutual information between $A$ and $B$. Then, for any (bounded) observables $O_A$ and $O_B$ of $A$ and $B$ respectively,
\be
\label{mutual}
I(A:B) \ge \frac{\left(\left<O_AO_B\right>-\left<O_A\right>\left<O_B\right>\right)^2}{2||O_A||^2||O_B||^2},
\ee            
where $\left<\cdots\right>$ is the expectation value of `$\cdots$' and $||\cdots||$ is the usual operator norm \cite{wolf}. Now, let $A$ and $B$ be some subregions of the conformal boundary of each exterior region at some constant time slice (but they do not necessarily have to be at the same time slice). Since $I(A:B)=S(A)+S(B)-S(A\cup B)$, when the minimal surface corresponding to $S(A\cup B)$ becomes the union of the minimal surfaces corresponding to $S(A)$ and $S(B)$, $I(A:B)$ vanishes at least at leading order in $1/N$. This is expected to happen at some point as the temperature decreases to zero, since the distance between any two points in the different exterior regions would diverge, as we mentioned above.\footnote{Since the minimal surface is defined by the real Lorentzian metric, this approach supports the assumption used above that the divergence of the real geodesic length between any two points in the different exterior regions implies cluster decomposition between them.} Then, by (\ref{mutual}), the two-sided correlator must factorize into a product of the one-sided correlators, up to subleading corrections in $1/N$. Important limitations of this approach compared to the above approach are that $O_A$ and $O_B$ cannot be a product of time-like seperated operators and that it is difficult to estimate the subleading correction to $I(A:B)$.\footnote{In general, a correction of order $N^0$ to $I(A:B)$ is expected to occur when $A$ and $B$ are seperated by a finite UV-regularized distance, but when the distance between $A$ and $B$ becomes infinite, such a correction, as well as higher order perturbative corrections (and non-perturbative ones in certain cases) in $1/N$, would probably vanish.} On the other hand, this approach complements the above approach by allowing $O_A$ and $O_B$ to be non-local operators such as Wilson loops. Also, this approach might be valid even if the bulk geometry is modified or not well-defined near or inside the horizon at low temperatures (for example, due to the mass gap of the black hole \cite{presk}, fuzzball-like structures,\footnote{However, we note that the fuzzball conjecture is usually associated with the microstates of a black hole, not their ensemble.} or the instability of the inner horizon), since the ``phase transition'' of the minimal surface for $S(A \cup B)$ could occur at much higher temperatures (see \cite{htmi} for a related discussion in case of the AdS Schwarzschild black hole).\footnote{Of course, if the subleading corrections turn out to be sensitive to the horizon geometry, this approach may also be affected by the potentially modified/ill-defined bulk geometry.}    

So far, we have dealt with correlators of boundary field theory operators. However, we can regard the bulk field operators as non-local operators in the boundary field theory \cite{bena,lowe1,lowe2,lowe3,lowe4,infall}, even at subleading orders in $1/N$ (i.e. when bulk fields are interacting) \cite{kabat,polchin}. By applying the zero temperature limit of (\ref{A}), (\ref{Abar}), and (\ref{AAbar}) to (\ref{bulkmain}), with $A$($\ov{A}$) being the non-local field theory operator corresponding to $\F^{\Rmnum{1}}$($\F^{\Rmnum{2}}$), we obtain exactly the same conclusion for bulk observables as for boundary observables.\footnote{One might think that the propagators and mode functions for the bulk fields in the microstate geometries may be significantly different from those in the black hole geometry and therefore the CFT construction of the bulk fields based on the black hole geometry may not be valid for the microstate geometries. However, if the bulk correlators are indistinguishable from those measured in the black hole geometry up to a non-perturbatively small difference, the effective geometry for a (low-energy) observer will be the black hole geometry.} Our conclusion may be incompatible with at least some versions of the fuzzball proposal, since different metric and field configurations corresponding to different microstates would probably be distinguished by measuring correlators of probe operators (whose number does not scale with $N$) to an accuracy of some power of $1/N$. There might be certain microstates in which the values of correlators differ from a universal value by some powers of $1/N$, but their number should be vanishingly small compared to $e^S$ when $N$ is large. 

Having presented our conclusion, let us now compare it with previous studies. In \cite{babel}, the authors argued that the probability of a large deviation from a universal value of probe correlators in microstates of the AdS Schwarzschild black hole in five dimensions is of order $e^{-S}\sim e^{-N^2}$, where $S$ is the black hole entropy. They also estimated the ratio of the variance to the squared mean of probe correlators to be of order $1/S$ based on calculations involving (grand)canonical ensembles of certain $1/2$ BPS states that correspond to bulk geometries without horizons, where $S$ is the entropy associated with the ensemble. Our results suggest that, at least for macroscopic extremal black holes, the deviation of any correlator in the black hole microstates from a universal value (i.e. the value of that correlator in the ensemble of the microstates) is non-pertubatively small in $N$ (perhaps exponentially small in $S$ or $N^2$). We believe that part of the reason they obtained such a large variance compared to ours is because they used (grand)canonical ensembles in which a temperature is introduced for the energy or conformal dimension of the BPS states (so the energy is allowed to fluctuate); in contrast, we essentially considered the variance in the microcanonical ensemble of the degenerate energy eigenstates. 

In \cite{typ}, it was derived that, if the variance of correlators in the ensemble of energy eigenstates in a certain energy range whose number is $e^S$ is of order 1, the variance of correlators in the ensemble of states spanned by these energy eigenstates is exponentially suppressed by a factor of $e^{-S}$. Our results shows that, for extremal black holes, there exists a natural basis of the black hole microstates, and the variance of the correlators in the former ensemble is not of order 1 but (exponentially) suppressed when $S$ is large. 
                          
In \cite{shige}, two-point correlators are computed in the microstates of the massless BTZ black hole at the orbifold point and compared with the correlators computed in the massless BTZ black hole geometry. While their results are not relevant to ours, since black holes whose horizon area is smaller than the Planck scale is involved in taking the zero temperature limit, it would be very interesting to extend their calculations to the (macroscopic) extremal BTZ black hole and see exactly what happens to correlators in its microstates.

Finally, we conclude with some open questions. The outstanding question is ``why and to what extent does the bulk cluster decomposition break down?'' One plausible answer is that subdominant geometries give rise to the violation of the cluster decomposition in the dominant geometry and therefore the size of the violation is of order $e^{-S}$. But, the question deserves a more definitive investigation. In certain cases, subdominant classical geometries may be difficult to explicitly find or not even exist; for instance, in case of canonical ensemble at fixed electric charge, the global AdS with a gas of charged particles is not a solution to the equations of motion and therefore cannot be a subdominant saddle \cite{emp2}. Also, a naive ``sum of semiclassical geometries'' does not always give the expected behavior for correlators from the point of view of the boundary field theory \cite{kleban}. Perhaps, the most serious problem with such an interpretation might be that, for (not toroidally compactified) planar black holes at finite $N$, their entropy is infinite and therefore the contribution from subdominant saddles naively vanishes. In this case, though, the number of degeneracy is infinite and therefore the zero variance of correlators does not necessarily imply that correlators in individual microstates should be all identical, which would be a contradiction. A related question is ``which correlators violate the bulk cluster decomposition and to what extent?'' In pariticular, it would be interesting to estimate the size of the violation as a function of the number of operators inside the correlator.      	             

Another important question is whether similar conclusions carry over to non-extremal black holes. For simplicity, let us consider a large AdS Schwarzschild black hole. Following essentially the same steps as in (\ref{n}) through (\ref{result}), we find that 
\be
\label{result2}
\left<|A-\left<A\right>|^2\right> = - {1\over Z}\sum_{i\not= j} e^{-{\b \over 2}(E_i + E_j)}|A_{ij}|^2 + \delta_{cor}, 
\ee
where $\delta_{cor} \equiv \langle \Psi | A\, \ov{A} | \Psi \rangle - \langle \Psi | A | \Psi \rangle \langle \Psi | \ov{A} | \Psi \rangle $, $\left<\cdots\right> \equiv 1/Z\sum_{i}e^{-\b E_i}\cdots$, and the sums are over all energy eigenstates. As before, $A$ and $\ov{A}$ can be any boundary field theory operators, including non-local ones representing the bulk field operators. Because there is no infinite throat for the Schwarzschild black hole, $\delta_{cor}$ is non-zero. If the LHS of (\ref{result2}) is ``small,'' the RHS has to be ``small'' as well, which would be a non-trivial fact about the AdS/CFT correspondence. However, in this case, it is not clear how ``small'' the LHS should be. For instance, in the high temperature regime, the thermal partition function
\be
Z(\b) = \tr e^{-\b H} = \int dE \Omega(E) e^{-\b E}
\ee 
is sharply peaked at an energy of order $N^2$ with a width of order $N$. This might suggest that the LHS of (\ref{result2}), which is a variance in canonical ensemble, could well be proportional to some powers of $1/N$. Perhaps, for such hot black holes, it might be more appropriate to consider microcanonical ensemble, as argued in \cite{typ} for example. 

\section*{Acknowledgments}
I would like to thank S. Kachru for support and help during the initial stages of this work and S. Shenker for valuable discussions. This work was supported by SLAC, which is operated by Stanford University for the US Department of Energy, under contract DE-AC02-76SF00515.

\end{document}